\documentclass[twocolumn,aps,prb,superscriptaddress,showpacs,amsmath,amssymb,floats]{revtex4}

\usepackage{amsmath}
\usepackage{amssymb}
\usepackage{graphicx}
\usepackage{keyval}
\usepackage{dcolumn}
\usepackage{bm}
\usepackage{color}
\usepackage{txfonts}
\usepackage[]{sidecap}

\begin{document}

\title{Interface-induced superconductivity and strain-dependent spin density wave in  FeSe/SrTiO$_3$ thin films}

\author{S. Y. Tan}
\affiliation{State Key Laboratory of Surface Physics, Department of Physics, and
Advanced Materials Laboratory, Fudan University, Shanghai 200433,
People's Republic of China}
\affiliation{Science and Technology on Surface Physics and Chemistry Laboratory, Mianyang 621907, Sichuan, People's Republic of China}

\author{M. Xia}
\author{Y. Zhang}
\author{Z. R. Ye}
\author{F. Chen}
\author{X. Xie}
\author{R. Peng}
\author{D. F. Xu}
\author{Q. Fan}
\author{H. C. Xu}
\author{J. Jiang}
\author{T. Zhang}

\affiliation{State Key Laboratory of Surface Physics, Department of Physics, and
Advanced Materials Laboratory, Fudan University, Shanghai 200433,
People's Republic of China}

\author{X. C. Lai}
\affiliation{Science and Technology on Surface Physics and Chemistry Laboratory, Mianyang 621907, Sichuan, People's Republic of China}

\author{T. Xiang}
\author{J. P. Hu}
\affiliation{Institute of Physics, Chinese Academy of Sciences, Beijing 100190, People's Republic of China}

\author{B. P. Xie}\email{bpxie@fudan.edu.cn}

\author{D. L. Feng}\email{dlfeng@fudan.edu.cn}

\affiliation{State Key Laboratory of Surface Physics, Department of Physics, and
Advanced Materials Laboratory, Fudan University, Shanghai 200433,
People's Republic of China}

\date{\today}

\begin{abstract}
\textbf{The record of superconducting transition temperature ($T_c$) has long been 56~K for the iron-based high temperature superconductors (Fe-HTS's). Recently, in single layer FeSe films grown on SrTiO$_3$ substrate, signs for a new 65~K $T_c$ record are reported. Here with \textit{in-situ} photoemission measurements, we substantiate the presence of spin density wave (SDW) in FeSe films,  a key ingredient of Fe-HTS that was missed in FeSe before, which weakens with increased thickness or reduced strain. We demonstrate that the superconductivity occurs when the electrons transferred from the oxygen-vacant substrate suppress the otherwise most pronounced SDW in single layer FeSe. Besides providing a comprehensive understanding of FeSe films and directions to further enhance its $T_c$, we establish the phase diagram of FeSe vs. lattice constant that contains all the essential physics of Fe-HTS's. With the simplest structure, cleanest composition and single tuning parameter, it is ideal for testing theories of Fe-HTS's.}
\end{abstract}

\maketitle

FeSe is the simplest and arguably most environmental-friendly Fe-HTS. The $T_c$ of bulk FeSe is only about 8~K (ref~\onlinecite{Wu}), however, it  reaches as high as 37~K under pressure \cite{FeSepressure}.  Theoretically, a collinear $2\times 1$ SDW order similar to that in the iron pnictides was predicted to be present in FeSe  (ref~\onlinecite{FeSeLDA}). However, partly due to the lack of high quality single crystals, such   SDW order was not found in FeSe or closely related compounds  before, and little is known about its electronic structure. Only spin fluctuations  around the SDW wave vector were found in Fe(Te,Se)  by inelastic neutron scattering\cite{Spinfluctuation}.  The magnetic structure in its closest sibling compound, FeTe, is bicollinear \cite{FeTe}, different from the SDW order observed in the iron pnictides.

Recently,  in single layer  FeSe thin films grown on   SrTiO$_3$ (STO) substrate by molecular beam epitaxy (MBE),  both Scanning Tunnelling Spectroscopy (STS) and angle-resolved photoemission spectroscopy (ARPES) experiments have observed the largest superconducting gap in Fe-HTS's \cite{FeSeXue,FeSeZhou}, which likely closes  above 65~K. Although further transport measurements  are needed to confirm whether the long standing 56~K record of $T_c$ is broken,  the remarkable properties of FeSe film, such as the role of the substrate and the superconducting mechanism, call for further exploration \cite{LeeFeSe}.

\begin{figure*}[t]
 \includegraphics[width=17cm]{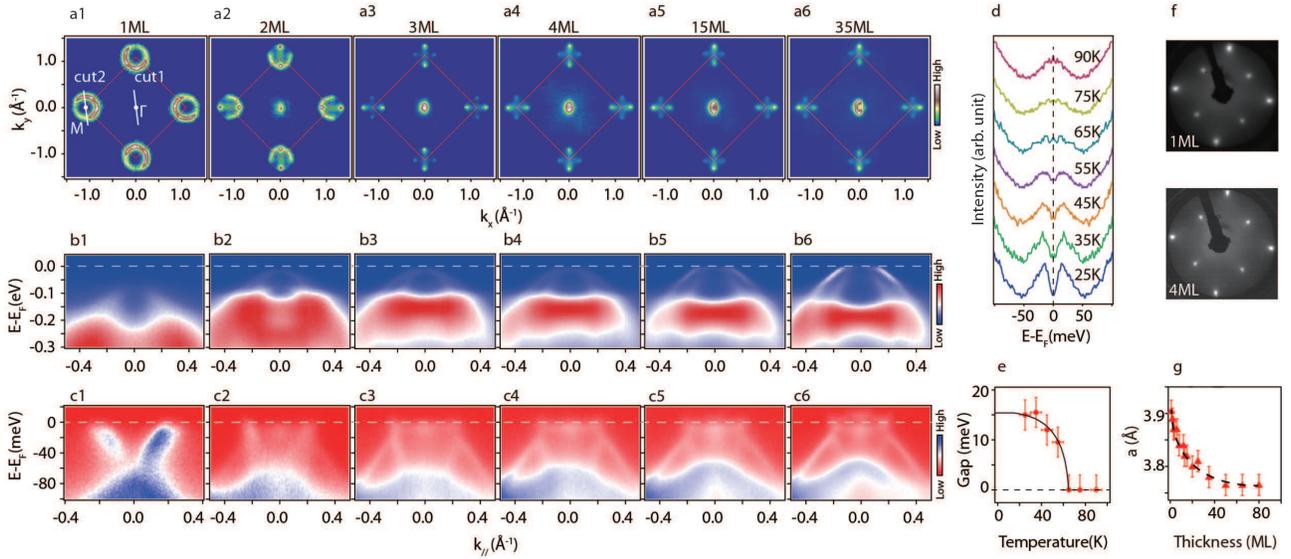}
\caption{\textbf{The electronic structure of FeSe  films as a function of thickness.} \textbf{a1-a6}, The thickness dependence of the Fermi surface as represented by the photoemission intensity at the Fermi energy at 30~K.  \textbf{b}, \textbf{c}, The thickness dependence of the band structure around (\textbf{b1-b6})  $(0,0)$ (cut \#1) and (\textbf{c1-c6}) $(\pi,0)$ (cut \#2) respectively. All data are taken at 30~K. \textbf{d}, Temperature dependence of the symmetrized EDC at the Fermi crossing of cut \#2 for 1~ML FeSe film, where the closing of the gap is shown by the filling up the states at $E_F$. \textbf{e}, The superconducting gap vs. temperature in 1~ML FeSe film. The gap is obtained following the standard fitting procedure described in ref. \onlinecite{ZhangKFeSe}. \textbf{f}, LEED (low energy electron diffraction) patterns for 1~ML and 4~ML FeSe film respectively.     \textbf{g}, The lattice constant $a$ of the top FeSe layer as a function of film thickness, which is derived based on the photoemission maps in panels \textbf{a1-a6}.   }
\label{general}
\end{figure*}

We have grown FeSe thin films  layer by layer on STO substrate with MBE as described in the Methods sections, and it was  transferred back and forth between the MBE chamber and  the ARPES chamber under ultra high vacuum for the continuous \textit{in-situ} growth and measurement of the electronic structure  as a function of thickness. The electronic structures of FeSe thin films  are presented in Fig.~\ref{general}. Since the photoemission intensity at the Fermi energy  ($E_F$)  reflects the Fermi surface, one found that the Fermi surface of the 1~ML (monolayer) FeSe film is  composed of four electron Fermi pockets at the zone corners  (Fig.~\ref{general}a1).  It is similar to the Fermi surface of the K$_x$Fe$_{2-y}$Se$_2$ (ref~\onlinecite{ZhangKFeSe}), except that the small electron pocket around $(0,~0,~\pi)$ in  K$_x$Fe$_{2-y}$Se$_2$ is absent here. Fig.~\ref{general}d shows the temperature dependence of the symmetrized photoemission spectrum taken at a Fermi crossing on the Fermi surface around M. The  measured  superconducting gap  ($\sim$ 15~meV) is larger than all the other  known bulk iron-based superconductors, and it closes  at a higher temperature of about  $65 \pm 5$~K (Fig.~\ref{general}e),  confirming the previous  \textit{ex-situ} ARPES measurement\cite{FeSeZhou}. This suggests  a new record of high   $T_c$ for Fe-HTS's, assuming it is not a pseudo-gap.

 \begin{figure}[b]
\includegraphics[width=8.5cm]{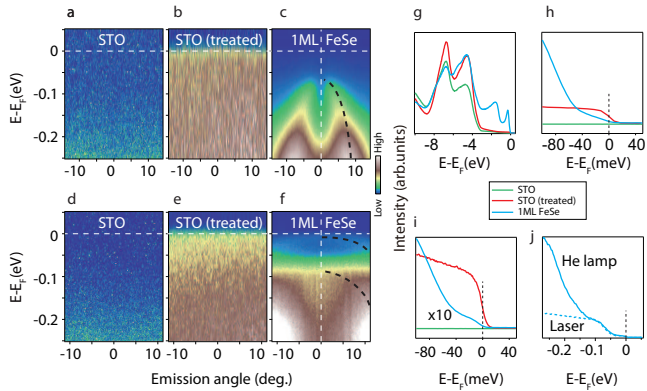}
\caption{\textbf{Electronic structure evolution during the growth of 1~ML    FeSe.} \textbf{a}, \textbf{b} and \textbf{c},  The photoemission intensity taken around normal emission for (\textbf{a}) the STO substrate after degassing at  550~$^{\circ}C$  for 3 hours, (\textbf{b}) STO substrate after 30~minutes heat treatment at 950~$^{\circ}C$ under Se-flux, and (\textbf{c})  1~ML FeSe respectively. Data were taken with 21.2~eV photons of a helium lamp. \textbf{d}, \textbf{e} and \textbf{f} are the same as panels \textbf{a}, \textbf{b} and \textbf{c}, except the data were taken with 7~eV photons  from  a  laser. \textbf{g}, and \textbf{h}, The comparison of (\textbf{g}) the valence band spectra and (\textbf{h}) the spectra near $E_F$ taken with a helium lamp in the above three cases.  \textbf{i}, The same as panel \textbf{h}  except the data were taken with a  7~eV laser. The 1~ML FeSe  data are amplifed ten times for clarity. \textbf{j}   compares the 1~ML FeSe data taken with laser and helium lamp, after normalized at -0.1~eV. All data were taken at 30K.
}
\label{1ML}
\end{figure}

Intriguingly, the Fermi surface topology changes dramatically for the 2~ML film (Fig.~\ref{general}a2). One not only observes  the Fermi surface of the 1~ML film, but also a new ``cross"-like Fermi surface at the zone corner. Some spectral weight also appears at the zone center. When the third FeSe layer is grown, the 1~ML Fermi surface is hardly detected. As shown in Figs.~\ref{general}a-\ref{general}c, the electronic structures of multilayer FeSe films are qualitatively similar, but with subtle differences. The spectral weight at the zone center is contributed by several hole-like bands. They do not cross $E_F$ in the 2~ML film; with increased thickness, they become stronger and cross $E_F$, and the hole pocket area increase slightly. The band structure near M is rather complicated, several bands cross  and give four small electron pockets that make the cross shape.  Such small pockets have been observed  in BaFe$_2$As$_2$ (ref~\onlinecite{LXYang}), EuFe$_2$As$_2$ (ref~\onlinecite{BoZhou}) in the SDW state. Low energy electron diffraction (LEED) pattern in Fig.~\ref{general}f shows that there is no noticeably lattice or charge superstructures, thus the complicated electronic structure is  not due to some  band folding by charge ordering.

In the photoemission intensity map of the 2~ML film (Fig.~\ref{general}a2), the center of the circular Fermi surface mismatches the center of the cross-shaped Fermi surface, indicating the different Brillouin zone sizes for the two FeSe layers, and thus different in-plane lattice constant $a$ along the Fe-Se-Fe direction. $a$ could be derived from the inversed Brillouin zone size determined by high symmetry points of photoemission maps, and we found that the lattice constant of the 1~ML film is severely expanded from the bulk value of 3.765~\AA ~to 3.905~\AA~ enforced by the STO lattice. With increased film thickness, $a$ relaxes rapidly (Fig.~\ref{general}g), and  reaches the bulk value at 35~ML and above.

The Fermi surface, band dispersion, and even superconducting gap  of the interfacial layer measured on the 2~ML film  are the same as those measured on the 1~ML  film (Figs.\ref{general}a2-\ref{general}c2), which indicates that the electronic structure of the interfacial FeSe layer is not affected by the surface FeSe layer. Therefore,  the interlayer coupling and charge transfer is very weak between them.  As an expected consequence of the weak coupling and its stoichiometry, the  top layer is charge neutral within 0.01$e^{-}$ per Fe accuracy  based on the calculated Luttinger volume.  Therefore, the   0.12 $e^{-}$ excessive electrons per Fe [derived from the Luttinger volume assuming the electron pockets are made of two degenerate ones like those of K$_x$Fe$_{2-y}$Se$_2$ (ref~\onlinecite{ZhangKFeSe})]   in the bottom FeSe layer is an interfacial effect.  Based on the LEED pattern in Fig.~\ref{general}f, the 1~ML film are of high quality, and there is no sign for any ordering of  Se vacancies, consistent with the  STM measurements \cite{FeSeXue}. Therefore the excessive electrons should come from the substrate.

\begin{figure*}[t]
 \includegraphics[width=13.5cm]{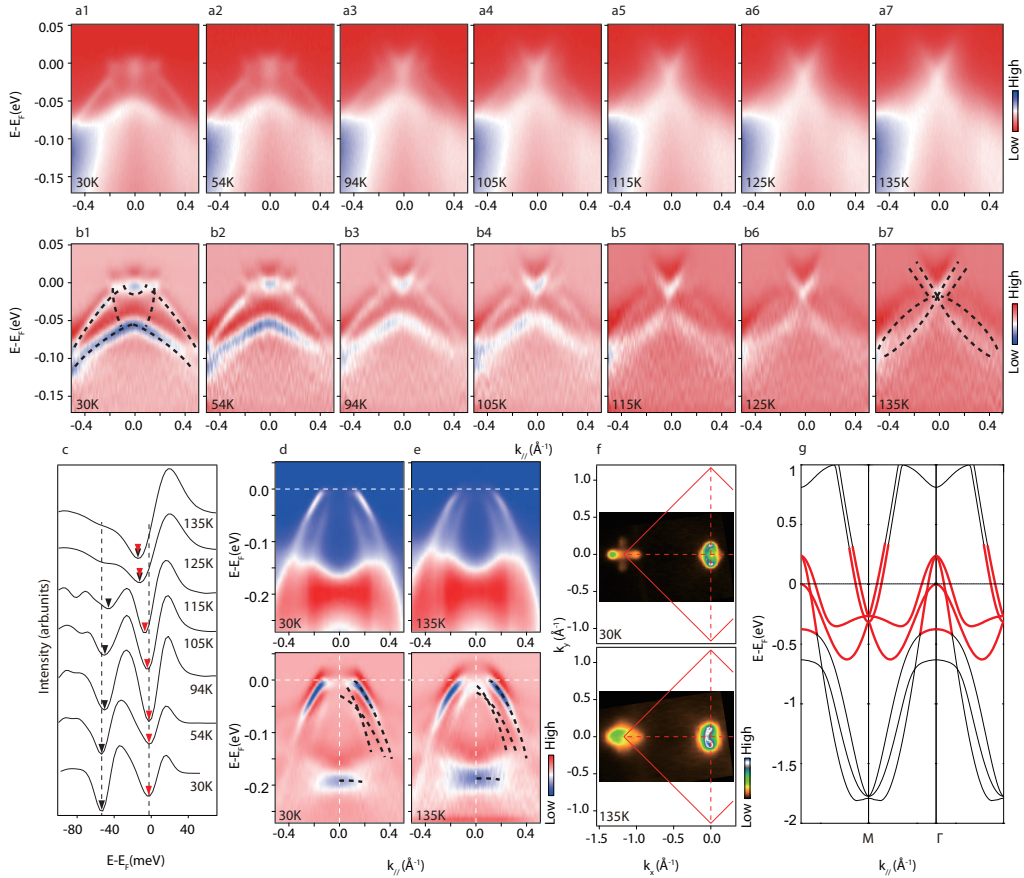}
\caption{  \textbf{Temperature dependence of the electronic structure for the 50~ML FeSe film.} \textbf{a} and \textbf{b},  (\textbf{a1-a7}) The photoemission intensity and (\textbf{b1-b7}) the corresponding second derivative with respect to energy around $(\pi,0)$. \textbf{c},  The  second derivative  of the EDC at $(\pi,0)$.  The  dip of the second derivative could  highlight the band position more clearly. \textbf{d}  and \textbf{e},  The electronic structure and the corresponding second derivative with respect to energy around $(0,0)$ at 30~K and 135~K respectively.   \textbf{f}, The photoemission intensity map at 30~K and 135~K respectively.  \textbf{g}, The calculated band structure of FeSe \cite{FeSeLDA}, the thickened part of the bands have been observed in our data.}
\label{Tdep}
\end{figure*}

To answer how  such charge transfer is induced at the interface \cite{LeeFeSe}, we have taken the photoemission data at the end of each of the three stages during the growth of 1~ML FeSe film (Fig.~\ref{1ML}). The  STO substrate was first degassed in ultra high vacuum  at  550~$^{\circ}C$  for 3 hours; then it was further  heat-treated at 950~$^{\circ}C$ under Se-flux for 30 minutes; and finally 1~ML FeSe was grown on the substrate. At the end of the first stage, the photoemission data are insulator-like,  no states are detected near $E_F$ as shown in Fig.~\ref{1ML}a (taken with 21.2~eV photons) and Fig.~\ref{1ML}d  (taken with 7~eV photons).  In the second stage, it is known that STO  loses oxygen at high temperatures  \cite{STO}, and metallic conductivity was observed for STO annealed in vacuum at 800~$^{\circ}C$ \cite{STO2}. The Se-flux acts as a cleaning agent during this stage, and no Se would be adsorbed at such high temperature. Metal-like Fermi step without momentum dependence is observed after the second stage (Figs.~\ref{1ML}b and \ref{1ML}e), mimicking the typical photoemission spectrum taken on  polycrystalline gold. The electrons in these states are most likely caused by the oxygen vacancies.  After the 1~ML FeSe was deposited, Figs.~\ref{1ML}c, and \ref{1ML}f show the typical dispersion of FeSe film.
 To identify the destiny of the non-dispersive STO states, Fig.~\ref{1ML}g compares the valence bands taken at normal emission after these three stages. The features around -4.5 and -6.7~eV are the STO valence states, which were taken  at the same condition and their intensities are comparable. Therefore, the photoelectrons from the substrate are not reduced noticeably by the top FeSe layer, thus the non-dispersive STO states would have been observed if they still exist after 1~ML FeSe is deposited. However, in Fig.~\ref{1ML}h, the Fermi step is clearly absent in the 1~ML data. The more bulk-sensitive photoemission data taken with 7~eV laser shows this contrast more clearly in Fig.~\ref{1ML}i.  As the charge has to be conserved, we conclude that the electrons in the localized oxygen-vacancy-induced states are transfered to the FeSe layer, and thus are responsible for the  electron doping in 1~ML FeSe.

A weak but dispersive band just below $E_F$ becomes visible in the laser data in Fig.~\ref{1ML}f,  and it  dominates the spectral weight in the first 40~meV below $E_F$.
Since compared with the Fermi step intensity of the oxygen vacancy states, the relative intensity of the weak dispersive feature is much weaker in the more bulk sensitive laser data than that in the Helium lamp data. Therefore, it can not be from STO.  In fact, it is visible in the Helium lamp data as well when the  spectra are normalized at low energies (Fig.~\ref{1ML}j), but was missed in the previous \textit{ex-situ} ARPES study\cite{FeSeZhou}.  Moreover, because its dispersion is different from the hole-like band in the 2~ML film, it should not be contributed by the appearance of small 2~ML FeSe regions. Judging from its intensity, it is likely made  of the  $d_{xy}$ orbital, whose  photoemission matrix element is weak near the zone center, while the stronger band at -0.1~eV is made of $d_{xz}/d_{yz}$ orbitals.

For the multi-layer FeSe films, the complicated electronic structure around the zone corner in Figs.~\ref{general}a2-a6, and ~\ref{general}c2-c6 is different from that of Fe(Te,Se) (ref~\onlinecite{FeTeSe}), but it is very similar to those observed in  BaFe$_2$As$_2$ (ref~\onlinecite{LXYang}) and NaFeAs (ref~\onlinecite{ChengHe}) in their SDW state. In fact, such dramatic electronic reconstruction has been proven to be the experimental hallmark of the  SDW or collinear antiferromagnetic order formation in all iron pnictides before \cite{LXYang,BoZhou,ChengHe,ZhangSrK,ZhangNa,ZhangFeTe,MingYi,MingYiNa}, and has been reproduced by dynamic mean field theory band calculations \cite{Yin1,Yin2}.  Such a electronic structure would disappear in the nonmagnetic state. To examine this,  Fig.~\ref{Tdep} presents the temperature dependence of electronic structure of the 50~ML film. Indeed, with increasing temperature, the separated bands gradually become degenerate again above 125~K (Figs.~\ref{Tdep}a and ~\ref{Tdep}b). Such a band separation has been shown to be caused by the different dispersions along the FM and AFM direction of the SDW order in  BaFe$_2$As$_2$ (ref~\onlinecite{LXYang}) and NaFeAs (ref~\onlinecite{ChengHe}).  Since photoemission is a very fast probe, it could sense the short-ranged nematic fluctuations or SDW that emerges at a higher characteristic temperature $T_A$ than the static ordering temperature $T_N$  observed by neutron scattering \cite{ChengHe,ZhangNa,ZhangFeTe,MingYiNa}. Therefore, our results prove that at least short-ranged SDW exists in the 50~ML FeSe thin films below 125~K.
Practically,  $T_A$ is the temperature that the separated bands become degenerate, and here it could be determined by the  merging of two dips in the second derivative of the EDC at $(\pi,0)$ in Fig.~\ref{Tdep}c.  On the other hand, we found that the band structure near zone center does not change much across the transition, and band folding due to SDW is not observed (Fig.~\ref{Tdep}d). The SDW-induced folding is often very weak for iron pnictides, which is likely caused by the lack of long range coherence.  Consequently, the Fermi surface is more strongly reconstructed near the zone corner (Fig.~\ref{Tdep}f).  At high temperatures, the  band structure of the 50~ML FeSe film  is determined from the second derivative of the photoemission data with respect to energy (Figs.~\ref{Tdep}e and \ref{Tdep}b7),  which shows a qualitative agreement with the calculation of bulk FeSe in the nonmagnetic state (Fig.~\ref{Tdep}g). The band renormalization factor is about  2$\sim$3, similar to those of most iron pnictides. Since high qualty  FeSe single crystal was not available, and particularly, the natural crystal surface is the (110) plane, it was very difficult to obtain the  electronic structure of bulk FeSe. Our data on the thick film give the first experimental bulk electronic structure  of FeSe.

\begin{figure}[t]
 \includegraphics[width=8.5cm]{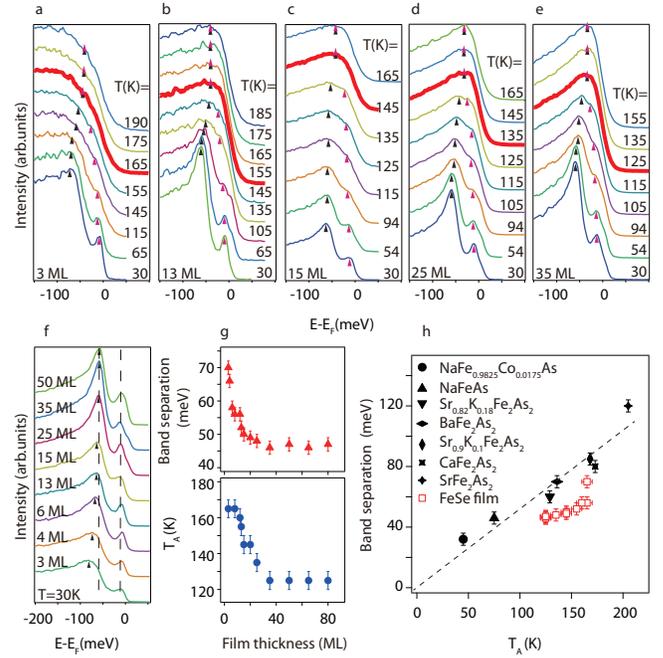}
\caption{ \textbf{Thickness dependence of the SDW behavior for  the multi-layer FeSe films.} \textbf{a-e}, Temperature evolution of the spectral lineshape at M for various thicknesses.  The two bands merge above a characteristic temperature $T_A$. \textbf{f}, The maximal band separation for various thickness. \textbf{g}, Band separation and $T_A$ as a function of thickness. \textbf{h}, Band separation as a function of  $T_A$. The dashed line is a guide for the eyes.}
\label{bandsplit}
\end{figure}

\begin{figure}[t]
\includegraphics[width=6.5cm]{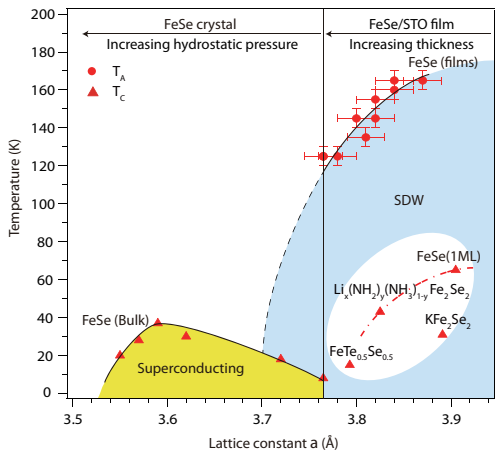}
\caption{   \textbf{Phase diagram of FeSe.} The $T_c$ and $T_A$ for FeSe  are ploted against the lattice constants. The right side is based on our thin film ARPES data, and the left side is based on the transport data of FeSe   single crystal under hydrostatic pressure taken from ref. \onlinecite{FeSepressure}.  The  dashed line represents the extrapolated  $T_A$'s, suggesting the existence of SDW order in  bulk FeSe under pressure. $T_c$ for other iron selenides are also plotted in the elliptical region.
}
\label{diagram}
\end{figure}

The band reconstructions observed in the 50~ML film  have been observed in all the films with more than 1~ML thickness (their data are presented in the supplementary Figs.~S1-S5). Using the temperature dependence of the EDC's at $(\pi,0)$ as representative, Fig.~\ref{bandsplit} examines how such signature of SDW order evolves with film thickness. As found similarly to Fig.~\ref{Tdep}c, the two features merge into one above the temperature $T_A$ in all cases  (Figs.~\ref{bandsplit}a-\ref{bandsplit}e). The band separation saturates at low temperatures, and Fig.~\ref{bandsplit}f collects EDC's at $(\pi,0)$ taken at 30~K. As summarized in Fig.~\ref{bandsplit}g,  the maximal separation decreases with increasing thickness, and asymptotically reaches a constant above 35~ML. Such separation characterizes the strength of the SDW order \cite{JuanJiangFeTe}, thus as expected, $T_A$ also decreases and becomes flat in the thick films. In Fig.~\ref{bandsplit}h, the relation between the maximal band separation and $T_A$ is shown together with all those existing ones measured on  the SDW state of various iron pnictides\cite{ZhangSrK,ZhangNa,MingYiBa,Ge}. The band separation follows  $T_A$ rather linearly in bulk samples. For FeSe films, the amplitude of  band separation is slightly smaller than  those of the iron pnictides, and there is a monotonic but nonlinear correlation  between $T_A$ and band separation. For many iron pnictides such as BaFe$_{2}$As$_{2}$ (refs \onlinecite{LXYang,ZhangSrK,MingYi}), $T_A$ and $T_N$ coincide with the structural transition temperature $T_S$ , however, for  compounds like NaFeAs (ref~\onlinecite{ChengHe}), $T_A>T_S>T_N$.   Since the $T_S$ is about 105~K for bulk FeSe (ref~\onlinecite{Wu}),  and $T_A$ of the thick films are about 125~K, we expect FeSe follows the latter case, namely, $T_N$ is  somewhere below $T_S$. This certainly  needs to be confirmed by more direct measurements in the future.

Our results in Fig.~\ref{Tdep} indicate that  the SDW order exists in thick FeSe films.  The $T_A$ determined for the thick films ($\sim$ 125~K)  agree well with the $\sim$130~K temperature scale determined by recent static and transient optical spectroscopies  on 460~nm thick (1,0,1) FeSe film grown on MgO substrate \cite{WuOptics}, where the phase transition nature was associated with nematicity-induced orbital or charge ordering  above the structural transition. Our data suggest that the optical data could be attributed to the SDW  fluctuations/ordering in FeSe. The large electronic structure reconstruction observed here explains the spectral weight transfer and partial gap opening in the optical data. In fact, the density functional theory calculations by Ma and coworkers \cite{FeSeLDA}  have shown that  the ground state of bulk FeSe is in the SDW state, instead of the bi-collinear antiferromagnetic  order as for FeTe, because the third nearest neighbor antiferromagnetic exchange interaction mediated by the $4p$ bands of Se is much smaller than that in  FeTe. Our results substantiate this prediction.

The asymptotic behavior in the thick films indicate that the film property is similar to its bulk behavior when the lattice is relaxed to its bulk value. Therefore the thickness dependence is actually a negative pressure (tensile strain) dependence of the FeSe system. In Fig.~\ref{diagram} we plot the phase diagram of FeSe as a function of lattice constant, together with the bulk FeSe phase diagram under hydrostatic pressure \cite{FeSepressure}, where the $T_c$ maximum is 37~K at 7~GPa.  The resulting phase diagram possesses all the same generic features as those  of  the iron pnictides, \textit{eg.}, superconductivity arises when the SDW is weakened, except the tuning parameter is lattice constant instead of doping. Particularly, it resembles the phase diagram of BaFe$_2$(As$_{1-x}$P$_x$)$_2$ (ref~\onlinecite{BaFeAsP}), Ba(Fe$_{1-x}$Ru$_x$)$_2$As$_2$ (ref~\onlinecite{BaFeRuAs}), where   physical or chemical pressure could induce similar  effects. Several other FeSe based superconductors have been plotted on the same phase diagram, K$_x$Fe$_{2-y}$Se$_2$ (ref~\onlinecite{ZhangKFeSe}) and Li$_x$(NH$_2$)$_y$(NH$_3$)$_{1-y}$Fe$_{2}$Se$_2$ (ref~\onlinecite{NH4FeSe})  are  heavily electron-doped, like the 1~ML FeSe/STO. Intriguingly, the electronic structure of the 1~ML FeSe film resembles that of  K$_x$Fe$_{2-y}$Se$_2$. We speculate that the $T_c$  of  K$_x$Fe$_{2-y}$Se$_2$ could have been much higher and closer to the dash-dotted line, if were not for its severe phase separation\cite{KFeSepressure,KFeSeFeiChen}.  The general trend here suggests that higher $T_c$ could be achieved in heavily electron doped  FeSe compound that has larger
lattice constant.

In the STM studies of the FeSe thin films\cite{FeSeXue}, it was surprising that the 1~ML FeSe is superconducting with high $T_c$, while the 2~ML FeSe layer appears ``semiconducting" with much reduced  density of states at $E_F$.  Besides the weak coupling between these two layers, the main reason is  the strong appearance of SDW order in the surface FeSe layer, which strongly
 reconstructs the electronic structure  and causes a suppression of spectral weight at $E_F$. Moreover, it eliminates  any proximity of the superconductivity from the interfacial superconducting FeSe layer.  Because the SDW order is enhanced with the increasing lattice constant in thinner films, it is reasonable to deduce that if the 1~ML FeSe were not so heavily doped by electrons transferred from the  substrate, it would have been in the strongest SDW state. In agreement with our results,  a recent  first-principal density functional calculations  show that  the ground state of 1~ML FeSe is a SDW state if it is undoped\cite{FeSefilmLDA}.  The same calculation also suggested that the 2~ML film is a slightly doped narrow gap semiconductor.

To summarize, we have shown that the property of FeSe is sensitive to the lattice, which provides a clean canonical system to study iron-based high superconductors.  We have provided compelling evidence for SDW order (at least short-ranged SDW order) in FeSe thin films and established the phase diagram of the FeSe system similar to the iron pnicides. The bulk FeSe electronic structure is revealed.  Our \textit{in situ} measurements not only confirm the possible $\sim$65~K   superconductivity in the 1~ML FeSe  film, but also show that the superconductivity of 1~ML FeSe is induced, when the SDW in it is suppressed  by the charge transferred from the oxygen vacancy states of the substrate. We suggest that one might further enhance $T_c$ by expanding the lattice and/or  optimizing the electron concentration via optimized substrate preparation. Moreover,  establishing the simplest possible and prototypical model system is often a critical step toward understanding complex phenomena in condensed matter physics; by identifying the missing SDW in FeSe and its evolution with lattice constant, we show that FeSe is the simplest/cleanest model system for studying Fe-HTS.

\textbf{Methods:}

FeSe thin film was grown on the TiO$_2$ terminated and Nb-doped SrTiO$_3$ (0.5\% wt) substrate with the molecular beam epitaxy (MBE) method following the previous report(ref~\onlinecite{FeSeXue}). The substrate was first ultra-sonically-cleaned by using deionized water, followed by drying in N$_2$ gas flow. This  produces a surface of strontium hydroxide. Then it was subsequently etched with buffered-oxide etchant  in order to remove the strontium hydroxide and form a TiO$_2$ terminated surface. After that, annealing treatment in air at 950~$^{\circ}C$ for 2 hours was carried out. Further cleaning by trichloroethylene  was needed to remove any dust before loading the substrate into the growth chamber with a base pressure of 5x10$^{-10}$ mbar.  In vacuum, the substrate was degassed at 550~$^{\circ}C$ for 3 hours, and then heated to 950~$^{\circ}C$ under the Se flux for 30 minutes. During growth, the substrate was kept at 490~$^{\circ}C$ with the selenium flux twenty times more than the Fe flux by co-deposition. The film thickness was monitored by crystal oscillator and confirmed by x-ray reflectivity measurements. After growth, the film was annealed at 600~$^{\circ}C$ for 3 hours, and directly transferred into the ARPES chamber with a typical vacuum of 1.2x 10$^{-11}$ mbar. ARPES was conducted with 21.2~eV photons from a helium discharge lamp, and a 7~eV laser. With a SCIENTA R4000 analyser,  energy resolutions of 6~meV (lamp) and 4~meV (laser) and an angular resolution of 0.3$^{\circ}$ were achieved. Aging effects were strictly monitored during the measurement.

\textbf{Acknowledgement:} We gratefully acknowledge  Prof. Qikun Xue, Xi Chen and Dr. Wei Li for sharing their thin film growth procedures, Prof. Zhongyi Lu for the file of  FeSe band structure, and the enlightening discussion with Prof. Chandra Varma. This work is supported in part by the National Science Foundation of China, and National Basic
Research Program of China (973 Program) under the grant Nos. 2012CB921400,
2011CB921802, 2011CBA00112, 2011CB309703.

\textbf{Author contributions:} S.Y.T., M.X.,  X.X., D.F.X., H.C.X., and R.P. built the MBE system and grew the films, S.Y.T., Y.Z., Z.R.Y., F.C., Q.F., J.J.
and B.P.X. performed ARPES measurements. S.Y.T., D.L.F., Y.Z., Z.R.Y., T.Z., T.X. and J.P.H. analyzed the ARPES data.  D.L.F. wrote
the paper. D.L.F. and X.C.L. are responsible for the infrastructure, project direction and planning.

\textbf{Additional Information:} The authors declare no competing
financial interests.  Correspondence and requests for materials
should be addressed to D.L.F. (dlfeng@fudan.edu.cn).

\end{document}